# Synergetic enhancement of power factor and suppression of lattice thermal conductivity via electronic structure modification and nanostructuring on Ni and B co-doped p-type Si-Ge alloy

Muthusamy Omprakash[1*], Saurabh Singh[1,3], Keisuke Hirata[1,3], Kentaro Kuga[1], Santhanakrishnan Harish[6], Masaru Shimomura[6], Masahiro Adachi[2], Yoshiyuki Yamamoto[2], Masaharu Matsunami[1], Tsunehiro Takeuchi[1,3,4,5]

1. Research Center for Smart Energy Technology, Toyota Technological Institute, Nagoya 468-8511, Japan.
2. Sumitomo Electric Industries Ltd., Hyogo 664-0016, Japan.
3. CREST, Japan Science and Technology Agency, Tokyo 102-0076, Japan.
4. Institute of Innovation for Future Society, Nagoya University, Nagoya 464-8603, Japan.
5. MIRAI, Japan Science and Technology Agency, Toyota Technological Institute, Tokyo 102-0076, Japan.
6. Graduate School of Science and Technology, Shizuoka University, Hamamatsu 432-8011, Japan.

*E-mail: omprakashmuthusamy@gmail.com

## Abstract

For simultaneously achieving the high-power factor and low lattice thermal conductivity of Si-Ge based thermoelectric materials, we employed, in this study, constructively modifying the electronic structure near the chemical potential and nano-structuring by low temperature and high-pressure sintering on nano-crystalline powders. Nickel was doped to create the impurity states near the edge of the valence band for enhancing the power factor with boron for tuning the carrier concentration. The nanostructured samples with the nominal composition of $Si_{0.65-x}Ge_{0.32}Ni_{0.03}B_x$ ($x = 0.01$, 0.02, 0.03, and 0.04) were synthesized by the mechanical alloying followed low-temperature and high-pressure sintering process. A large magnitude of Seebeck coefficient reaching 321 $\mu VK^{-1}$ together with a small electrical resistivity of 4.49 m$\Omega$cm, leads to a large power factor of 2.3 $Wm^{-1}K^{-2}$ at 1000 K. With successfully reduced thermal conductivity down to 1.47 $Wm^{-1}K^{-1}$, a large value of *ZT* ~1.56 was obtained for $Si_{0.65-x}Ge_{0.32}Ni_{0.03}B_{0.03}$ at 1000 K.

# 1. Introduction

Thermoelectric devices made of solid-state materials allow us to directly converting waste heat to useful electricity. Such devices involving no moving components are capable of working longer time than other devices working with mechanical parts. Besides, no emission of global warming gases during the generation of electricity makes thermoelectric devices considered as one of the environmentally friendly technologies of power generation.

The efficiency of energy conversion in thermoelectric devices is an increasing function of dimensionless figure of merit ($ZT$), which is determined from the physical properties of constituent materials [1]. The expression of $ZT$ in terms of physical quantities is given by the following equation.

$$ZT = S^2 \sigma T / (\kappa_l + \kappa_e) \qquad (1)$$

Here, $S$, $\sigma$, $\kappa_l$, and $\kappa_e$ are Seebeck coefficient, electrical conductivity, lattice thermal conductivity, and electron thermal conductivity, respectively. The product of squared Seebeck coefficient and electrical conductivity, which exist in the numerator of Eq. (1), is known as power factor ($PF = S^2 \sigma$) and is often used as another measure of thermoelectric performance. The materials used in the thermoelectric generator of high-efficiency in energy conversion should possess a large power factor together with low thermal conductivity. Besides, it is very challenging to improve $ZT$ because of above parameters are interrelated.

However, many strategies have been proposed for enhancing the power factor of thermoelectric materials. Those are tuning of the carrier concentration, band engineering, resonant states, convergence of electronic bands, quantum confinement, and energy filtering effect [2-6]. A decrease in the '$\kappa_l$' without degrading the power factor has been also widely tried to achieve a large $ZT$ through alloying, nanostructuring, and grain boundary interfaces [7-14].

Aforementioned strategies described above have led to large magnitude of $ZT$ exceeding unity for the various materials such as PbTe [15], AgPb$_{(m)}$SbTe$_{(2+m)}$ [16], Bi$_2$Te$_3$ [17], Bi$_2$Te$_3$ / Sb$_2$Te$_3$ superlattice [18], Bi-Sb-Te [19]. However, those materials consist of expensive and /or toxic elements such as Pb, Te, Sb, and these constituent elements prevented us from making the wide commercial application.

In sharp contrast to those thermoelectric materials, Si-Ge is widely considered a cheap, non-toxic, thermally stable material, and has potential for various applications working at high temperatures above 500 °C. Even though the Si-Ge thermoelectric materials having the slightly smaller $ZT$ value of 0.9 and 0.5 for the *n* and *p*-type materials, thermoelectric generators consisting of Si-Ge alloys are currently used for a deep-space mission to stably convert radioisotope heat into electricity [20].

It would be very important to know that, the Si-Ge alloy has the difference in the mean free path of phonon and electrons about 200-300 nm and ~ 5 nm [21], respectively. Hence, the nanostructuring approach in the scale range of 10-100 nm has less influence on electron transport properties but a major contribution to reduce '$\kappa_l$'. As a consequence, many researchers reported higher $ZT$ value, for example, Joshi *et al.* improved 90% of $ZT$ (up to 0.95), which is almost double as compared with *p*-type Si-Ge material currently in space missions ($ZT$ = 0.5). Such a drastic improvement in $ZT$ was mainly attributed to the scattering of phonon at the nano-grain interfaces [22]. S. Bathula *et al.* reported the high $ZT$ of 1.2 at 1173 K was achieved by creating the nano to mesoscale dimension defects in the *p*-type Si-Ge [23]. Basu *et al.* reported an improvement of *ZT* to 1.84 at 1073 K in the *n*-type Si-Ge nanostructure on the achievement of low thermal conductivity about 0.9 Wm$^{-1}$K$^{-1}$ [24]. Another approach named nanocomposites was developed to improve the *ZT* which was proposed by Bergman and Levy in 1991 [25]. N. Mingo *et al.* discussed the semiclassical calculation of electron and phonon transport, which suggests the size of metallic nanoparticles, which are distributed from 3 – 30 nm in the Si-Ge matrix could minimize the thermal conductivity [26]. In addition, thermoelectric properties of composite materials, for example, the SiC nanoparticles in the *n*-type Si-Ge matrix having the lowest '*κ*' of 1.9 Wm$^{-1}$K$^{-1}$ was observed and leads to high *ZT* about 1.7 at 1173 K [27]. The YSi$_2$ included in the *p*-type Si-Ge improved the *ZT* of 1.81 at 1100 K which was due to the presence of low '*κ$_l$*' [28].

Here, our strategy for enhancing the *ZT* value of thermoelectric materials consists of two parts: (a) constructively modifying the electronic structure together with tuning carrier concentration and (b) reducing of '*κ*' by employing nano-structuring. Using this strategy, recently, we succeeded in developing *p*-type Si-Ge showing a high *ZT* of 1.38 and 1.63 at 973 K for the thin film and bulk nano-composites [29-31]. This large value of *ZT* = 1.63 rather a low temperature was brought about by a low thermal conductivity of 1.3 Wm$^{-1}$K$^{-1}$ and a high power-factor of 2.1 mWm$^{-1}$K$^{-2}$ which were caused by the nano-structuring, the electronic structure modification with a small amount of gold (Au), and optimization of carrier concentration with boron (B).

More recently, this strategy was also applied for an *n*-type Si-Ge alloy, in which nano-structuring with Fe impurity was employed with P-doping to realize the highest *ZT* of 1.88 at 873 K [32]. This result suggests that Fe 3d rather than Au 5d would have a more preferable effect in electronic structure. Therefore, in this study, we tried to increase the *ZT* of *p*-type Si-Ge by employing 3d transition metal elements as the source of impurity states near the valence band edge in the electronic density of states and reduce '$\kappa_l$' by nanostructuring.

## 2. Calculation and Experimental procedures

We calculated the electronic structure of Ge with various transition metal element substitutions, by using density functional theory based code implemented in the WIEN2K (Full-potential Linearized Augmented Plane Wave method) program [33]. The Generalized Gradient Approximation in the form of Perdue-Burke-Ernzerhof (PBE) was used for the exchang correlation function [34]. The *k*-mesh size in the irreducible edge of the first Brillouin zone was set to be 8×8×8 for the Self-consistency field (SCF) and density of states calculations. The energy convergence criteria were fixed to the $10^{-5}$ Ry for total energy/cell.

To obtain the information about impurity states due to transition metal (TM) elements, we also performed the cluster calculations using a simulation package known as DVX-*α*, in which a linear combination of the atomic orbital method was used with the X*α* potential for electron correlation [35]. The supercell of a size of 2×2×2 (64 atoms) was made by using the conventional Ge unit cell (8 atoms), and one of the Ge atoms in the supercell is replaced by transition metal elements (Ni, Cu, Au). The partial density of states results was obtained by atomic clusters calculation consisting of the 71 atoms ($Ge_{34}TM_1H_{36}$) centered at the X atom from relaxed structure, a similar way reported in our previous work on the Al-Mn-Si system [36].

Results from the above mentioned electronic structure calculations suggested that Ni is one of the most plausible elements for improving *PF* of Si-Ge thermoelectric material. The detailed results will be explained later.

The samples possessing nominal composition of $Si_{0.65-x}Ge_{0.32}Ni_{0.03}B_x$ ($x$ = 0.01, 0.02, 0.03, and 0.04) were prepared by employing mechanical alloying followed by low-temperature sintering at 973 K under high pressure of 400 MPa for a rather long soaking time of 4 h. The high purity Ni (4N) and Ge grains (5N) were melted using the arc furnace in a pressurized Ar gas atmosphere. Afterward, the Ge-Ni mother ingot was crushed into powder and sealed with the high purity Si (5N) and boron powder in a Zirconia container of 40 ml in volume together with Zirconia balls of 6 mm in diameter under pressurized Ar gas atmosphere in a glove box. The mixed powders were milled by using the high-energy planetary ball mill (Fritsch-P7) at a rotation speed of 700 rpm for 6 h with a pause time of 10 min for every 1 h milling to avoid the re-welding of powder.

The synthesized powders were collected from the container in the glove box filled with 1 atm Ar gas to avoid oxidation. The consolidated powders were loaded into hardened stainless steel with an internal diameter of 10 mm and subsequently sintered using a spark plasma sintering system (SPS) with high pressure of 400 MPa at 973 K for 4 h, respectively. The synthesized samples of nominal composition $Si_{0.65-x}Ge_{0.32}Ni_{0.03}B_x$ ($x$ = 0.01, 0.02, 0.03, and 0.04) were denoted as $x$ = 0.01, $x$ = 0.02, $x$ = 0.03, and $x$ = 0.04 throughout the manuscript.

The crystalline phases involved in the samples were identified by conventional powder x-ray diffraction (XRD) with Cu-Kα radiation using a BRUKER D8 Advance. The microstructure, nanoparticle distribution, and compositions were analyzed by scanning electron microscope (SEM) using a JEOL JSM-6330F and energy-dispersive x-ray spectroscopy (EDX) using a JED-2140GS with the accelerating voltage of 20 kV, respectively. The distribution of the nanocrystalline, defects by lattice distortion, and nano-size precipitation were confirmed by TEM analysis (JEOL JEM-1400). The density of the bulk sample was evaluated by using the Archimedes method.

The thermal diffusivity was investigated in the wide temperature range from 300 to 1000 K in argon flow atmosphere by using a laser flash method in LFA457 manufactured by Netzsch. The specific heat capacity was evaluated by a comparative method using Pyroceram as a standard sample. The specific heat capacity was also cross-checked by differential scanning calorimetry using RIGAKU DSC 8231 system. The thermal conductivity ($\kappa$) was estimated by the product of thermal diffusivity ($\alpha$), specific heat capacity ($Cp$), and density ($d$).

The sound velocity was measured by using an ultrasonic pulse generator (Olympus). The sound velocity was calculated by using the relationship $C = 2l / \Delta t$ where $C$, $l$, and $\Delta t$ are the sound velocity, the thickness of a sample, and the time difference between two waves, respectively.

The carrier concentrations were roughly estimated from the Hall coefficient determined by using the Physical Properties Measurement System (PPMS) developed by Quantum Design, Inc. The Hall coefficient ($R_H$) was measured by four probe configurations with a magnetic field up to 9 T. The carrier concentration was roughly calculated by using the relation $n_H = -R_H^{-1}/e/^{-1}$, where $e$ represents the unit charge of the electron.

The electrical resistivity and Seebeck coefficient were measured using a homemade system with the conventional four-probe technique and steady-state method, respectively, in the temperature range of 300 to 1000 K under a vacuum atmosphere.

Seebeck coefficient of the standard constantan bar was measured to calibrate the measurement setup, which is found to stay within 6% deviation from the absolute measured value (38.65 μVK$^{-1}$ at 300 K). The figure of merit was evaluated from the measured electrical resistivity, Seebeck coefficient, and thermal conductivity in the temperature range of 300 to 1000 K.

The photoemission spectroscopy was measured using MBS A-1SYS hemispherical photoelectron spectrometer and a monochromatized He lamp at room temperature. The energy resolution was set to about 25 meV. Clean sample surfaces were obtained *in situ* by fracturing under the pressure of $4 \times 10^{-8}$ Pa. The chemical potential was estimated by fitting the Fermi-Dirac distribution function of the Fermi edge in an evaporated gold measured before and after the sample measurements.

## 3 Results

We calculated the electronic density of states for Ge with various transition metals substitutions such as Ni, Au, and Cu. Figure 1 shows the partial density of states (PDOS) of Ge containing one of the selected elements, Ni, Au, and Cu. Among them, the Ni element produces a sharp peak near the edge of the valence band. It means that a large Seebeck coefficient might be obtained with metallic electrical conduction. As a consequence of this rough analysis of the electronic structure of Ge containing a small amount of transition metal element, we selected Ni as one of the most plausible elements potentially leading to a significant enhancement in power factor.

The XRD patterns of un-milled and 6 h milled samples were shown in Fig. 2. All samples were identified as a single phase of the diamond structure with space group Fd-3m with background having humps around 28° and 50°. The background shape indicates the presence of an amorphous phase. The average crystallite size was estimated to be 8 ~ 9 nm for all the samples from the analyses using Scherer's formula ($Dp$)

$$Dp = (0.94\lambda) / (\beta \cos \Theta), \qquad (2)$$

where, $Dp$, $\lambda$, $\beta$, and $\Theta$, are the averaged size of crystals, X-ray wavelength, line broadening, and Bragg angle, respectively.

The crystallization temperature of the amorphous phase was investigated by DSC analysis as shown in Fig. 3. An exothermic peak was exhibited at 856 K for the sample $x = 0.03$, which was lower than that of un-doped Si-Ge alloy at 1015 K [37]. In this case, lowering the crystallization temperature would be attributed to the Ni substitution, and this phenomenon is known as metal-induced crystallization [38]. Subsequently, the milled powders were consolidated and sintered at 1000 K with high pressure of 400 MPa for 4h.

Figure. 4 (a) shows the XRD pattern of sintered samples. The sharpened peaks indicate that the crystal growth occurred during the sintering process. The estimated grain size was found to be 26~29 nm. The XRD peaks were found to be shifted towards lower angles with increasing B concentration $x$. The peak shift was shown in Fig. 4 (b) with the (111) peak as a representative, and the corresponding lattice constant was plotted in Fig. 4 (c). It shows that the lattice constant was increased with increasing B, which indicated that B was soluted in the diamond structure Si-Ge matrix. This result is consistent with the previously reported B doped (0.25 – 2.0 at%) $Si_{0.8}Ge_{0.2}$ bulk sample [39]. The grain size, lattice constant, and unit cell volume were summarized in Table 1.

Very tiny impurity peaks were observed for all the samples at 25.2°, 26.8°, 31.3°, 41.8°, and 44.6°, which were ascribed to the peaks from $Ge_2Ni$, $GeO_2$, $GeNi_{1.5}$. The formation of the tiny amount of these impurities was agreed well with the very tiny solubility limit of Ni in Ge. Unfavorable oxidization was also confirmed by the presence of $GeO_2$, despite that its amount was very small. The experimentally determined density, theoretical density, relative density, and crystallite size of sintered samples were summarized in Table. 2. All the samples were possessed in relative density of more than 91.5%.

Figure 5 shows the secondary electron SEM images obtained for the polished surface of the sintered samples. All the samples have distributions of isolated nanometer-sized particles (< 100 nm), micro-grains (1 ~ 10 μm) made up of the aggregated nano-particles, and the very small amount of $Ge_2Ni$ and $GeNi_{1.5}$ precipitations. The precipitation of $Ge_2Ni$ and $GeNi_{1.5}$ was indicated by the solid black arrow line in Fig. 5.

The EDX mapping analysis clearly shows the homogeneous distribution of Si, Ge, B, and precipitation of Ge-Ni as shown in Fig. 6 (a-d). Table. 3 shows the quantitative detail of the elemental composition obtained from the electron probe microanalysis (EPMA). It revealed that the Ge composition was reduced from the nominal concentration and this reduction was naturally attributed to the precipitation of $Ge_2Ni$, $GeO_2$, and $GeNi_{1.5}$ compounds. The oxygen concentration was also detected in the Si-Ge matrix about 3.13 at.%, 2.69 at.%, 2.68 at.%, and 4.6 at.% for the samples of $x = 0.01$, 0.02, 0.03, and 0.04, respectively. The formation of a tiny amount of $SiO_2$ might occur at the surface of nano-grains probably due to the milling process and short time exposure of sample powder while loading it into the SPS chamber after the ball milling step.

The TEM image of the $x = 0.03$ sample after 6 h milling is shown in Fig. 7 (a). The Si-Ge nanocrystals with ~8 nm in averaged diameter were observed for the as-milled sample. The disordered structure in the lattice, which corresponds to the amorphous phase, was also observable in Fig.7 (a). The disordering was most likely caused by the high energy impact of balls and powders in the milling container during high-speed ball milling. Figure 7 (b) shows the TEM images of the sintered sample, in which the nanocrystalline with size 5 ~ 25 nm and Ge-Ni precipitation were observed. The distribution of nanocrystalline Si-Ge and Ge-Ni precipitates was highlighted by the white color solid and dotted line. The obtained grain size was well consistent with the XRD results.

Noteworthy, the SEM and TEM results indicated that the sizes of the hierarchical nanoparticles < 100 nm, micro-grain boundaries approximately from 1~ 10 μm, and the lattice disorder were fabricated by low-temperature with high-pressure sintering process and these characteristics should be adequate to scatter the short, mid, and long-wavelength phonons.

### 3.4 Thermoelectric properties

Figure 8 (a) shows the temperature dependence of electrical resistivity measured from 300 to 1000 K. All the samples possess a rather weak temperature dependence, which is typical behavior of metallic conduction in heavily disordered materials. The positive temperature coefficient of electrical resistivity (TCR) is observed for all the samples, and it means that the mean free path of electrons is slightly shortened with increasing temperature due to the phonon excitations but its variation is very small because the highly disordered structure made the electrons very close to the strongest scattering limit known as the Ioffe-Regel-Mott limit [40] In such a case, electrical resistivity is determined mainly by the density of states averaged over a few $k_\mathrm{B}T$ in energy width centered at the chemical potential.

The electronic density of states averaged over a few $k_\mathrm{B}T$ in energy width centered at chemical potential is also reflected in the Hall coefficient. In Fig.8 (b) and (c), the room temperature electrical resistivity and Hall coefficient were plotted as a function of nominal boron concentration. We also plotted electrical resistivity as a function of the Hall coefficient in Fig. 8 (d). We confirmed from these plots that the structure quality of $x = 0.04$ is worse than other samples most likely due to both the secondary phases and the oxidation.

The Seebeck coefficient plotted in Fig. 8 (e) shows a positive sign for all the samples, which indicates that the holes are the majority in carriers, i.e., *p*-type material. The Seebeck coefficient monotonically increased with increasing temperature regardless of B concentrations, and the samples were naturally considered as degenerate semiconductors. The sample obtained at $x = 0.01$ exhibits the largest value of Seebeck coefficient of 372 µVK$^{-1}$ at 937 K. Notably, the Seebeck coefficient was higher than previously reported *p*-type Si-Ge+YSi$_2$ [28], Si-Ge+SiMo [41], Si-Ge+TiB$_2$ [42], and Si-Ge+CrSi$_2$ [43] as shown in Table 6. It possessed a small decrease in magnitude at a high temperature above 950 K. This decrease would be accounted for with the electron-hole excitation, which is often explained in terms of the "*bi-polar diffusion effect*".

For the samples of $0.02 \leq x \leq 0.04$ at high temperatures above 700 K, the magnitude of the Seebeck coefficient decreased with increasing *x*, while it possesses almost the same magnitudes regardless of $x$ at the temperature range below 700 K. This behavior cannot be explained in terms of a single parabolic band model with varying carrier concentration, and therefore we need to consider some variations, which was presumably caused by Ni, in the electronic density of states near the band edge of the valence band.

The power factor ($PF = S^2 \sigma$) was plotted as a function of temperature from 300 to 1000 K in Fig. 8 (f). The power factor increased with increasing temperature. The largest magnitude of power factor $PF = 2.3$ Wm$^{-1}$K$^{-2}$ was observed at 1000 K for the sample $x = 0.03$. The magnitude of $PF$ significantly decreased with increasing or decreasing $x$ from $x = 0.03$ regardless of temperature.

In Fig. 9 (a), the thermal conductivity was plotted as a function of temperature in the range from 300 to 1000 K. The decreasing behavior of thermal conductivity with increasing temperature was observed in all the samples, which was attributed to the intensified phonon-phonon scattering with increasing temperature. The decreasing ratio with temperature was rather weak regardless of B concentration. This tendency is safely attributed to the disordering structure and densely persisting grain boundaries that cause temperature-independent phonon scattering.

As a consequence, the thermal conductivity was decreased significantly to 1.2 Wm$^{-1}$K$^{-1}$ at 973 K for $x = 0.01$. Importantly, obtained thermal conductivity of all samples at 1000 K was lower than that of previously reported Si-Ge+YSi$_2$ [28], Si-Ge+SiMo [41], Si-Ge+TiB$_2$ [42], and Si-Ge+CrSi$_2$ [43] samples. With increasing $x$, the magnitude of thermal conductivity showed a slightly but finite increase over the whole temperature range, and this tendency becomes more obvious with increasing $x$. Since we already realized that carrier concentration increased with increasing B (values shown in table.4), this tendency is attributed to the increases in electron thermal conductivity.

The electron thermal conductivity '$\kappa_e$' was roughly estimated by employing Widemann-Franz law $\kappa_e = \sigma L_0 T$, where $\sigma$, $L_0$, and $T$ represent the electrical conductivity, Lorentz number ($L_0 = 2.45 \times 10^{-8}$ W$\Omega$K$^{-2}$), and absolute temperature, respectively. The '$\kappa_e$' was increased with increasing the temperature due to heat carried by the charge carriers as shown in Fig. 9 (b). Moreover, the carrier concentration dependence of '$\kappa_e$' was also clearly observed for all the samples over the entire temperature range from 300 to 1000 K.

The lattice thermal conductivity '$\kappa_l$' was also roughly estimated by eliminating $\kappa_e$ from the measured '$\kappa$' as $\kappa_l = \kappa - \kappa_e$ and results were plotted as a function of temperature in Fig. 9 (c). It reveals that the low lattice thermal conductivity '$\kappa_l$' was nearly constant for about 1 Wm$^{-1}$K$^{-1}$ for $x = 0.02$ and $x = 0.03$. This mainly comes from the scattering of a wide range of phonons by the variable scattering centers such as nano to micro grain boundaries, defects in the lattices, and tiny precipitation. Whereas, for $x = 0.02$ and $x = 0.04$, the estimated '$\kappa_l$' was increased up to 1.2 Wm$^{-1}$K$^{-1}$ at high temperature. We consider that this increase was not attributed to that of '$\kappa_l$' but of '$\kappa_{el}$' in association with the bipolar effect, because the bipolar effect is generally enhanced in high temperatures and not involved in the Widemann-Franz law. Since the bandgap in Si-Ge alloys tends to decrease with Ge concentration, the evolution of the bipolar effect would be brought about by the tiny decrease of Si composition.

Figure. 9 (d) show the dimensionless figure of merit $ZT$ as a function of temperature in the range from 300 to 1000 K. The magnitude of $ZT$ increased with increasing temperature regardless of B concentrations. It possessed strong B concentration dependence with showing the maximum value of $ZT$ = 1.56 at $x$ = 0.03, 1000 K. The sample obtained at $x$ = 0.02 possessed slightly smaller magnitude of $ZT$ than that in $x$ = 0.03, but the value of $ZT$ both of $x$ = 0.01 and 0.04 became less than half of the values observed for $x$ = 0.03. The largest $ZT$ of $x$ = 0.03 was brought about by a large power factor of 2.3 mWm$^{-1}$K$^{-2}$ together with low thermal conductivity of 1.47 Wm$^{-1}$K$^{-1}$ at 1000 K. It would be worth noting that the obtained $ZT$ of sample $x$ = 0.03 was higher than that of previously reported Si-Ge+SiMo [41], Si-Ge+TiB$_2$ [42], and Si-Ge+CrSi$_2$ [43]. It was lower than that of Si-Ge +YSi$_2$ [28] because of higher electrical resistivity of 4.49 mΩcm as compared to Si-Ge+YSi$_2$ resistivity of about 2.1 m Ωcm at 1000 K.

## 4. Discussion

To discuss the role of B on the transport properties behavior at high temperature in more detail, all the TE property data are plotted at 873 K as a function of nominal boron concentration.

Figure 10 (a) shows the electrical resistivity as a function of nominal boron concentration at 873 K. The electrical resistivity decreased with increasing the boron concentration, which was attributed to increasing the carrier concentration. Whereas, the resistivity was increased for the sample $x$ = 0.04, although it consists of a high value in the roughly estimated carrier concentration 7.2 x 10$^{19}$ cm$^{-3}$. The unusual increase in electrical resistivity at $x$ = 0.04 would be accounted for with the slightly larger oxygen concentration and /or the presence of precipitated impurity phases. Noteworthy, Schierning *et al.* investigated the influence of oxygen on microstructure and thermoelectric properties of Si nanocomposites [44]. Miura *et al.*, synthesized Si nanocrystalline by gas-phase method and exposed it to air for several hours [45]. In both cases, the electrical resistivity of the bulk sample is increasing with exposure time due to the formation of a $SiO_2$ layer on Si nanoparticles. Therefore, reduction of oxygen concentration would be the next target to increases in *ZT* of nano-structured Si-Ge containing Ni and B.

The Seebeck coefficient of sample $x = 0.02$, $x = 0.03$, and $x = 0.04$ possessed almost constant value regardless of $x$ over a wide temperature range from 300 to 700 K as highlighted in Fig. 10 (b). This behavior cannot be accounted for with the single parabolic band model and can be one of the weak evidence of Ni-induced electronic structure modification near the band edge. The linear decrease of the Seebeck coefficient with increasing $x$, on the other hand, was observed at 873 K as shown in Fig 10 (c). This would be related to the "*bipolar effect*" as that also became obvious in the thermal conductivity at high temperatures. Figure10 (d) shows the highest power factor value of about 2 $mWm^{-1}K^{-2}$ was observed for $x = 0.03$ at 873 K.

To investigate impurity states near the valence band edge, photoemission spectroscopy was employed at 300 K using He II in the high vacuum condition $10^{-7}$ Pa. Figure 11 shows the valence electron spectrum of all samples measured at an excitation energy of 40 eV. It shows that there was no significant evidence of impurity state near the valence band edge although PDOS calculation showed a sharp peak. Since the amount of Ni was only 3 at.%, and therefore we considered that the impurity states would be too small to be recognized.

For more reasonable analysis, we investigated the transport properties of the bulk $Si_{0.62}Ge_{0.35}B_{0.03}$ and compared them with the results of $Si_{0.62}Ge_{0.32}B_{0.03}Ni_{0.03}$ to study the effect of Ni substitution on the transport properties. Figure 12 (a) shows that electrical resistivity as a function of temperature in the range from 300 to 1000 K. It depicts that the electrical resistivity of the $Si_{0.62}Ge_{0.35}B_{0.03}$ sample showed rather significant negative TCR most likely due to hopping conduction occurred by the localized carriers in the disordered sample while $Si_{0.62}Ge_{0.32}B_{0.03}Ni_{0.03}$ shows the weak positive TCR due to the delocalized carriers under the nearly strongest scattering limit in the disordered sample. Moreover, the electrical resistivity of $Si_{0.62}Ge_{0.32}B_{0.03}Ni_{0.03}$ was lower than that of the sample $Si_{0.62}Ge_{0.35}B_{0.03}$ over the entire temperature range from 300 to 1000 K. These experimental facts indicate that substitution of Ni also increasing the carrier concentration.

The Seebeck coefficient of sample $Si_{0.62}Ge_{0.32}B_{0.03}Ni_{0.03}$ was lower than that of $Si_{0.62}Ge_{0.35}B_{0.03}$ over the entire temperature range due to low electrical resistivity influenced by co-doped Ni and B. The Seebeck coefficient decreased at 910 K because bipolar diffusion occurred for the sample $Si_{0.62}Ge_{0.35}B_{0.03}$ as compared with sample $Si_{0.62}Ge_{0.32}B_{0.03}Ni_{0.03}$ as shown in Fig. 12 (b). Hence, co-doped Ni and B help to decrease the electrical resistivity rather than increasing the Seebeck coefficient. The suppression of the Seebeck coefficient might be associated with the neither absence of impurity states near the valence band nor higher carrier concentration. However, the maximum power factor was obtained for $Si_{0.62}Ge_{0.32}B_{0.03}Ni_{0.03}$ than that of sample $Si_{0.62}Ge_{0.35}B_{0.03}$ due to low electrical resistivity together with moderate Seebeck coefficient as shown in Fig. 12 (c).

In general, most of the heat carried by the lattice thermal conductivity '$\kappa_l$' and acoustic phonon is mainly contributed for the '$\kappa_l$'. The average mean free path (MFP) of phonon was roughly estimated by using the relation $l = Vg\tau$, where, $l$, $Vg$, and $\tau$ represent phonon mean free path, sound velocity, and relaxation time, respectively. We assume that the relaxation time of all samples (τ) about $0.88 \times 10^{-12}$ s, which was theoretically predicted for the disorder $Si_{0.8}Ge_{0.2}$ [46]. The average sound velocity of longitudinal, transverse, and MFP was estimated at 300 K and is summarized in table 5. Figure 13 shows the average MFP as a function of nominal boron concentration. It shows that the average MFP of all samples was closely matched with the theoretically reported value of 4.1 nm [45].

Hence, low lattice thermal conductivity '$\kappa_l$' about ~1 Wm$^{-1}$K$^{-1}$ was observed for $x$ = 0.01, 0.02, and 0.03 at 873 K as shown in Fig. 14 (a). It implies that the sample possessed sufficient scattering centers such as nanoparticles, micro grain boundaries, defects in the lattice, and tiny precipitation to scatter the various wavelength of phonons. Besides, lattice thermal conductivity '$\kappa_l$' was lower than the previously reported Si-Ge+YSi$_2$ [28], Si-Ge+SiMo [41], Si-Ge+TiB$_2$ [42], and Si-Ge+CrSi$_2$ nanocomposites [43] as shown in Fig. 14 (b). The scattering process of various wavelengths of phonons is schematically illustrated in Fig. 14 (c).

Figure. 15 (a) shows the maximum *ZT* value of 1.56 was achieved for the sample $x$ = 0.03 due to the large power factor 2.3 mWm$^{-1}$K$^{-2}$ together with low thermal conductivity of 1.47 Wm$^{-1}$K$^{-1}$ at 1000 K, respectively. The maximum *ZT* of sample $x$ = 0.03 was higher than the other sample and previously reported Si-Ge+SiMo [41], Si-Ge+TiB$_2$ [42], and Si-Ge+CrSi$_2$ nanocomposites [43] as shown in Fig. 13 (b). As discussed above, oxygen influences higher electrical resistivity. Thus, the key factor to further increase *ZT* is reducing the oxygen in the sample.

## 5. Conclusions

The nanostructured samples were synthesized by mechanical alloying, and low-temperature with a high-pressure sintering process at the composition of $Si_{0.65-x}Ge_{0.32}Ni_{0.03}B_x$ ($x$ = 0.01, 0.02, 0.03, and 0.04). The low lattice thermal conductivity '$\kappa_l$' of 0.91 Wm$^{-1}$K$^{-1}$ was achieved for the sample $x$ = 0.03 due to scattering of various wavelength phonons at the grain boundaries size in the range of 5 nm – 10 μm, tiny Ge-Ni precipitation, and dislocation in the lattice. The impurity state in association with Ni presumably contributed to the increase of the Seebeck coefficient and decrease of electrical resistivity. Thus, a larger power factor together with low thermal conductivity 1.47 Wm$^{-1}$K$^{-1}$ allowed to reach a high *ZT* of 1.56 at 1000 K in the optimized composition $Si_{0.62}Ge_{0.32}Ni_{0.03}B_{0.03}$.

**References**

[1] *CRC Handbook of Thermoelectrics*; Rowe, D. M., Ed.; CRC Press:Boca Raton, 1995.

[2] K. F. Hsu, S. Loo, F. Guo, W. Chen, J. S. Dyck, C. Uher, T. Hogan, E. K. Polychroniadis, M. G. Kanatzidis, Cubic AgPb(m)SbTe(2+m): bulk thermoelectric materials with high figure of merit, Science 303 (2004) 818, https://doi.org/10.1126/science.1092963.

[3] J. P. Heremans, V. Jovovic, E. S. Toberer, A. Saramat, K. Kurosaki, A. Charoenphakdee, S. Yamanaka, G. J. Snyder, Enhancement of thermoelectric efficiency in PbTe by distortion of the electronic density of states, Science 321 (2008) 554, https://doi.org/ 10.1126/science.1159725.

[4] B. Paul, P. K. Rawat, P. Banerji, Dramatic enhancement of thermoelectric power factor in PbTe:Cr co-doped with iodine, Appl. Phys. Lett. 98 (2011) 262101

[5] Y. Pei, X. Shi, A. LaLonde, H. Wang, L. Chen, G. J. Snyder, Convergence of

electronic bands for high performance bulk thermoelectrics, Nature 473 (2011) 66, https://doi.org/10.1038/nature09996

[6] S. V. Faleev, F. Leonard, Theory of enhancement of thermoelectric properties of materials with nanoinclusions, Phys. Rev. B 77 (2008) 214304, https://doi.org/10.1103/PhysRevB.77.214304

[7] Snyder, G. J.; Toberer, Complex thermoelectric materials, Nat. Mater. 7 (2008) 105, https://doi.org/10.1038/nmat2090

[8] W. Li, L. Zheng, B. Ge, S. Lin, X. Zhang, Z. Chen, Y. Chang, Y. Pei, Promoting SnTe as an Eco-Friendly Solution for p-PbTe Thermoelectric via Band Convergence and Interstitial Defects, Adv. Mater. 29 (2017) 1605887, https://doi.org/10.1002/adma.201605887

[9] Rogers, L. M. Valence band structure of SnTe, J. Phys. D: Appl. Phys.1 (1968) 845, https://doi.org/10.1088/0022-3727/1/7/304

[10] W. Liu, X. Tan, K. Yin, H. Liu, X. Tang, J. Shi, Q. Zhang, C. Uher, Convergence of conduction bands as a means of enhancing thermoelectric performance of n-type $Mg_2Si_{(1-x)}Sn_{(x)}$ solid solutions, Phys. Rev. Lett. 108 (2012) 166601, https://doi.org/10.1103/PhysRevLett.108.166601

[11] X. Liu, T. Zhu, H. Wang, L. Hu, H. Xie, G. Jiang, G. J. Snyder, X. Zhao, Low Electron Scattering Potentials in High Performance $Mg_2Si_{0.45}Sn_{0.55}$ Based Thermoelectric Solid Solutions with BandConvergence, Adv. Energy Mater. 3 (2013) 1238, https://doi.org/10.1002/aenm.201300174

[12] Y. Tang, Z. M. Gibbs, L. A. Agapito, G. Li, H. S. Kim, M. B. Nardelli, S. Curtarolo, G. J. Snyder, Convergence of multivalleybands as the electronic origin of high thermoelectricperformance in $CoSb_3$ skutterudites. Nat. Mater. 14 (2015) 1223, https://doi.org/10.1038/nmat4430

[13] S. Roychowdhury, R. Panigrahi, S. Perumal, K. Biswas, Ultrahigh Thermoelectric Figure of Merit and Enhanced Mechanical Stability of p-type $AgSb_{1-x}Zn_xTe_2$. ACS Energy Lett. 2 (2017) 349, https://doi.org/10.1021/acsenergylett.6b00639

[14] V. Jovovic, J. P. Heremans, Measurements of the energy band gap and valence band structure of AgSbTe$_2$, Phys. Rev. B: Condens. Matter Mater. Phys. 77 (2008) 245204, https://doi.org/10.1103/PhysRevB.77.245204

[15] Y. Pei, J. Lensch-Falk, E. S. Toberer, D. L. Medlin, G. J. Snyder, High thermoelectric performance in PbTe due to large nanoscale Ag2Te precipitates and la doping, Adv. Funct. Mater. 21 (2011) 241, https://doi.org/10.1002/adfm.201000878

[16] E. Quarez, K.-F. Hsu, R. Pcionek, N. Frangis, E. K. Polychroniadis, M. G. Kanatzidis, Nanostructuring, Compositional Fluctuations, and Atomic Ordering in the Thermoelectric Materials AgPb$_m$SbTe$_{2+m}$. The Myth of Solid Solutions, J. Am. Chem. Soc. 127 (2005) 9177, https://doi.org/10.1021/ja051653o

[17] M. Michiardi, I. Aguilera, M. Bianchi, V. E. de Carvalho, L. O. Ladeira, N. G. Teixeira, E. A. Soares, C. Friedrich, S. Blügel, P. Hofmann, Bulk band structure of Bi$_2$Te$_3$, Phys. Rev. B: Condens. Matter Mater. Phys. 90 (2014) 075105, https://doi.org/10.1103/PhysRevB.90.075105

[18] R. Venkatasubramanian, E. Siivola, T. Colpitts, B. O'Quinn, Thin-Film Thermoelectric Devices with High Room- Temperature Figures of Merit, Nature 413 (2001) 597, https://doi.org/10.1038/35098012

[19] J. Li, Q. Tan, J. Li, D. Liu, F. Li, Z. Li, M. Zou, K. Wang, BiSbTe-Based Nanocomposites with High ZT: The Effect of SiC Nanodispersion on Thermoelectric Properties. Adv. Funct. Mater. 23 (2013) 4317, https://doi.org/10.1002/adfm.201300146

[20] G.L. Bennett, J.J. Lombardo, and B.J. Rock, U.S. Radioisotope Thermoelectric Generator Space Operating Experience, The nuclear Engineer, 25 (1984) 49.

[21] Y.C. Lan, A.J. Minnich, G. Chen, Z.F. Ren, Enhancement of Thermoelectric Figure-of-Merit by a Bulk Nanostructuring Approach, Adv. Funct. Mater. 20 (2010) 357, https://doi.org/10.1002/adfm.200901512.

[22] G. Joshi, H. Lee, Y. Lan, X. Wang, G. Zhu, D. Wang, R. W Gould, D. C. Cuff, M. Y. Tang, and M. S. Dresselhaus, Enhanced thermoelectric figure-of-merit in nanostructured p-type silicon germanium bulk alloys, Nano Lett. 8 (2008) 4670, https://doi.org/10.1021/nl8026795

[23] S. Bathula, M. Jayasimhadri, B. Gahtori, N. Kumar Singh, A.K. Srivastava, A. Dhar, The role of nanoscale defects features in enhancing the thermoelectric performance of p-type nanostructured SiGe alloy, Nanoscale. 7 (2015) 12474. https://doi.org/10.1039/C5NR01786F

[24] R. Basu, S. Bhattacharya, R. Bhatt, M. Roy, S. Ahmad, A. Singh, M. Navaneethan, Y. Hayakawa, D.K. Aswal, S.K. Gupta, Improved thermoelectric performance of hot pressed nanostructured n-type SiGe bulk alloys, J. Mater. Chem. A 2 (2014) 6922, https://doi.org/10.1039/C3TA14259K

[25] D. J. Bergman and O. Levy, Thermoelectric properties of a composite medium, J. Appl. Phys. 70 (1991) 6821, https://doi.org/10.1063/1.349830

[26] N. Mingo, D. Hauser, N. P. Kobayashi, M. Plissonnier, and A. Shakouri, "Nanoparticle-in-Alloy" Approach to Efficient Thermoelectrics: Silicides in SiGe, Nano Lett. 9 (2009) 711, https://doi.org/10.1021/nl8031982

[27] S. Bathula, M. Jayasimhadri, B. Gahtori, A. Kumar, A.K. Srivastava, A. Dhar, Enhancement in thermoelectric performance of SiGe nanoalloys dispersed with SiC nanoparticles, Phys.Chem. Chem. Phys. 19 (2017) 25180, https://doi.org/10.1039/C7CP04240J

[28] S. Ahmad, A. Singh, A. Bohra, R. Basu, S. Bhattacharya, R. Bhatt, K.N. Meshram, M. Roy, S.K. Sarkar, Y. Hayakawa, A.K. Debnath, D.K. Aswal, S.K. Gupta, Boosting thermoelectric performance of p-type SiGe alloys through in-situ metallic YSi$_2$ nanoinclusion, Nano Energy 27 (2016) 282, https://doi.org/10.1016/j.nanoen.2016.07.002

[29] M. Adachi, S. Nishino, K. Hirose, M. Kiyama, Y. Yamamoto, and T. Takeuchi, High Dimensionless Figure of Merit $ZT$= 1.38 Achieved in p-Type Si-Ge-Au-B Thin Film, Materials Transaction, 61 (2019) 1014, https://doi.org/10.2320/matertrans.MT-M2019310

[30] M. Omprakash, K. Delime-Codrin, G. Swapnil, S. Saurabh, S. Nishino, M. Adachi,

Y. Yamamoto, M. Matsunami, S. Harish, M. Shimomura, T. Takeuchi. Au and B co-doped p-type Si-Ge nanocomposites possessing ZT = 1.63 synthesized by ball milling and low-temperature sintering, Jpn. J. Appl. Phys. 58 (2019) 125501, https://doi.org/10.7567/1347-4065/ab4fb9

[31] M. Omprakash, S. Swapnil, S. Singh, K. Delime-codrin, S. Nishino, M. Adachi, Y. Yamamoto, M. Matsunami, S. Harish, M. Shimomura, and T. Takeuchi, Enhancement of the Thermoelectric Performance of Si-Ge Nanocomposites Containing a Small Amount of Au and Optimization of Boron Doping, Journal of Electronic Materials, 49 (2020) 2813, https://doi.org/10.1007/s11664-019-07857-5

[32] K. Delime-Codrin, M. Omprakash, S. Ghodke, R. Sobota, M. Adachi, M. Kiyama, T. Matsuura, Y. Yamamoto, M. Matsunami, and T. Takeuchi, Large figure of merit $ZT$ = 1.88 at 873 K achieved with nanostructured $Si_{0.55}Ge_{0.35}$ ($P_{0.10}Fe_{0.01}$) Appl. Phys. Express 12 (2019) 045507, https://doi.org/10.7567/1882-0786/ab08b7

[33] P. Blaha, K. Schwarz, G. K. H. Madsen D. Kvasnicka and J. Luitz, WIEN2k: an Augmented Plane Wave Plus Local Orbitals Program for Calculating Crystal Properties Austria: Karlheinz Schwarz Technische University at Wien, 2001.

[34] J. P. Perdew, S. Burkey and M. Ernzerhof Phys. Generalized Gradient Approximation Made simple, Phys. Rev. Lett. 77 (1996) 3865, https://doi.org/10.1103/PhysRevLett.77.3865

[35] H. Adachi, M. Tsukada, and C. Satoko, Discrete Variational Xα Cluster Calculation. I. Application to Metal Clusters, J. Phys. Soc. Jpn. 45 (1978) 875, https://doi.org/10.1143/JPSJ.45.875

[36] A. Yamamoto, H. Miyazaki, M. Inukai, Y. Nishino, T. Takeuchi, Thermoelectric properties of Al-Mn-Si C40 phase containing small amount of W or Ta, Jpn. J. Appl. Phys. 54 (2015) 071801, https://doi.org/10.7567/JJAP.54.071801

[37] M. Omprakash, S. Nishino, S. Ghodke, M. Inukai, R. Sobota, M. Adachi, M. Kiyama, Y.Yamamoto, T. Takeuchi, H. Santhanakrishnan, H. Ikeda, and Y. Hayakawa, Low Thermal Conductivity of Bulk Amorphous $Si_{1-x}Ge_x$ Containing Nano-Sized Crystalline Particles Synthesized by Ball-Milling Process, Journal of Electronic

Materials, 47 (2018) 3260, https://doi.org/10.1007/s11664-018-6103-2

[38] N. Vouroutzis1, J. Stoemenos1, N. Frangis1, G. Z. Radnóczi, D. Knez, F. Hofer, and B. Pécz, Structural characterization of poly-Si Films crystallized by Ni Metal Induced Lateral Crystallization, Scientific Reports, 9 (2019) 2844, https://doi.org/10.1038/s41598-019-39503-9

[39] R. Murugasami, P. Vivekanandhan, S. Kumaran , R. Suresh Kumar , T. John Tharakan, Simultaneous enhancement in thermoelectric performance and mechanical stability of p-type SiGe alloy doped with Boron prepared by mechanical alloying and spark plasma sintering, J. Alloys Compd. 773 (2019) 752. https://doi.org/10.1016/j.jallcom.2018.09.029

[40] A.F. Ioffe and A.R. Regel, Non-crystalline, amorphous and liquid electronic semiconductor, Prog. Semicond. 4 (1960) 237.

[41] Y. Li, J. Han, Q. Xiang, C. Zhang, and J. Li, Enhancing thermoelectric properties of p-type SiGe by SiMo addition, J. Mater. Sci. Mater. Electr. 30 (2019) 9163, https://doi.org/10.1007/s10854-019-01245-9

[42] S. Ahmad, R. Basu, P. Sarkar, A. Singh, A. Bohra, S.Bhattacharya, R. Bhatt, K.N. Meshram, S. Samanta, P.Bhatt, M. Navaneethan, Y. Hayakawa, A.K. Debnath, S.K. Gupta, D.K. Aswal, K.P. Muthe, and S.C. Gadkari, Enhanced thermoelectric figure-of-merit of p-type SiGe through TiO2 nanoinclusions and modulation doping of boron, Materialia 4 (2018) 147, https://doi.org/10.1016/j.mtla.2018.09.029

[43] Z. Zamanipour and D. Vashaee, Comparison of thermoelectric properties of p-type nanostructured bulk $Si_{0.8}Ge_{0.2}$ alloy with $Si_{0.8}Ge_{0.2}$ composites embedded with $CrSi_2$ nano-inclusions, J. Appl. Phys. 112 (2012) 093714, https://doi.org/10.1063/1.4764919

[44] G. Schierning, R. Theissmann, N. Stein, N. Petermann, A. Becker, M. Engenhorst, V. Kessler, M. Geller, A. Beckel, H. Wiggers, and R. Schmechel, Role of oxygen on microstructure and thermoelectric properties of silicon composites, J. Appl. Phys. 110, 113515 (2011), https://doi.org/10.1063/1.3658021

[45] A. Miura, S. Zhou, T. Nozaki, and J. Shiomi, Crystalline-Amorphous Silicon Nanocomposites with Reduced Thermal conductivity for Bulk Thermoelectrics, Appl. Mater. Interfaces 7 (2015) 13484, https://doi.org/10.1021/acsami.5b02537

[46] P. Norouzzadeh, A. Nozariasbmarz, J. S. Krasinski, and D. Vashaee, Thermal conductivity of nanostructured $Si_xGe_{1-x}$ in amorphous limit by molecular dynamics simulation, J. Appl. Phys. 117, 214303 (2015).

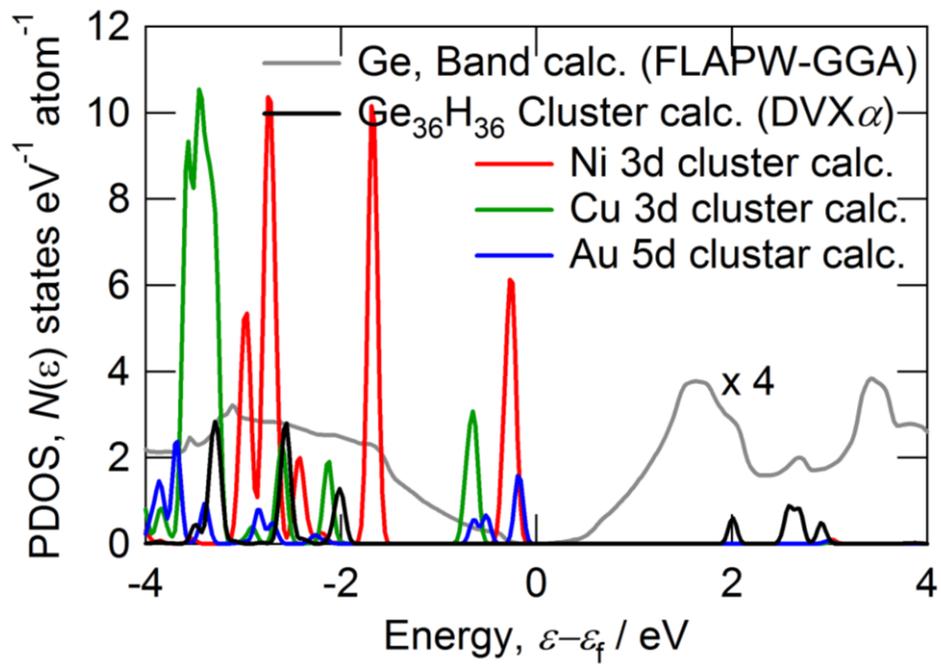

Fig. 1 Partial DOS of Ge containing various transition metal Ni, Au, and Cu.

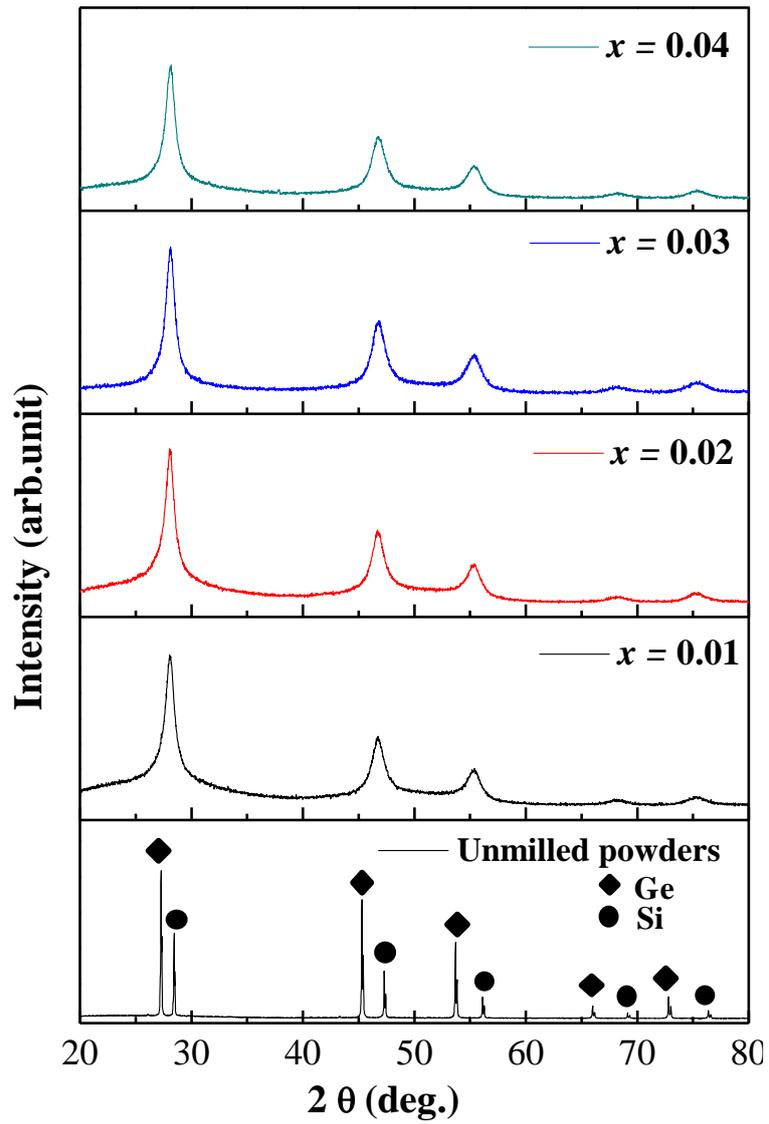

Fig. 2 XRD patterns of sample $x = 0.01$, $x = 0.02$, $x = 0.03$, and $x = 0.04$ after 6 h milling at 700 rpm.

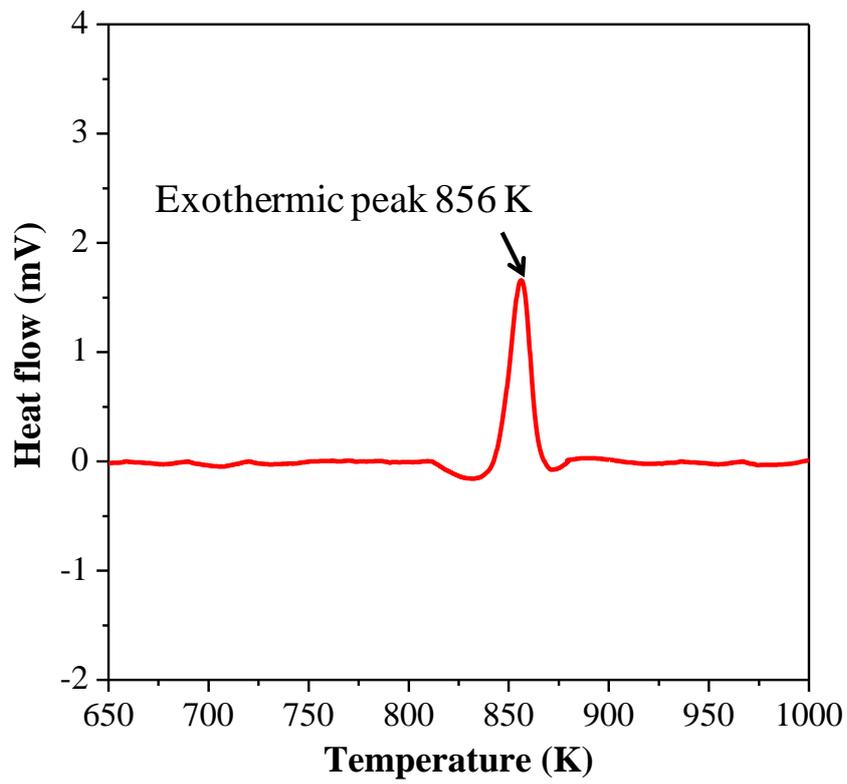

Fig. 3 Thermal stability analysis of 6 h milled sample $x = 0.03$ by DSC analysis

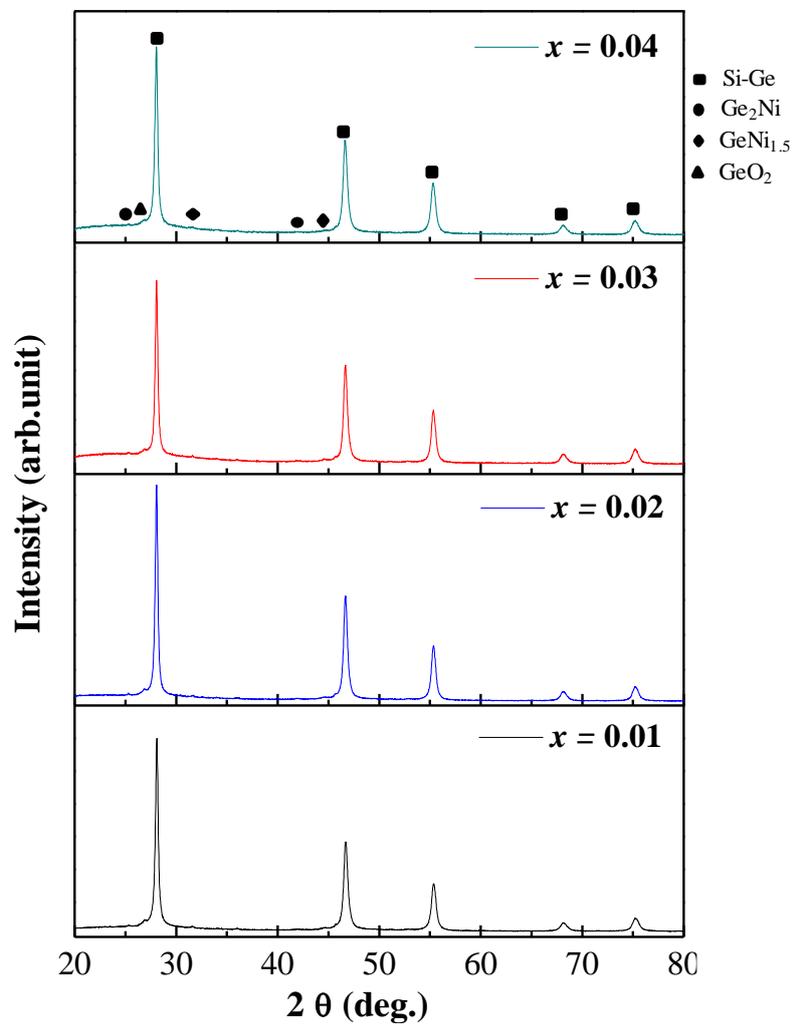

Fig. 4 a

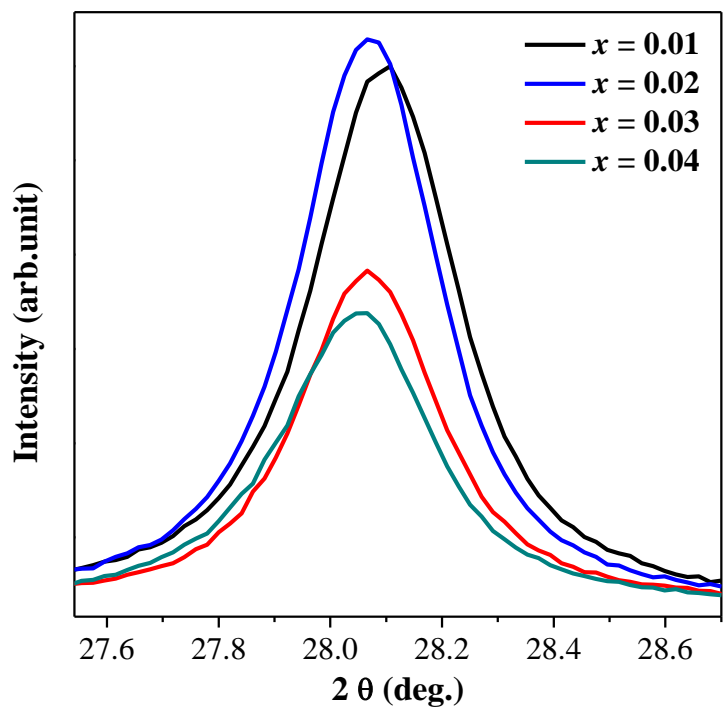

Fig. 4 b

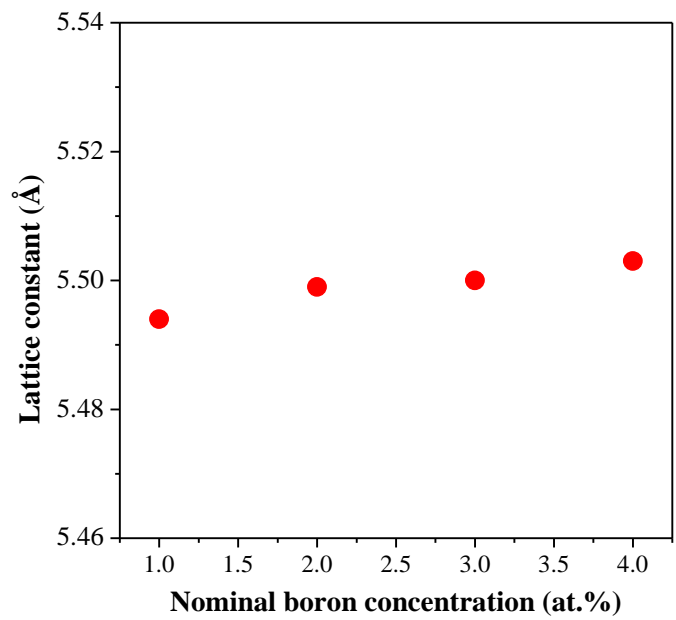

Fig. 4 c

Fig. 4 (a) XRD pattern of sample $x = 0.01$, $x = 0.02$, $x = 0.03$, and $x = 0.04$ after sintering, (b) enlarged XRD pattern, and (c) lattice constant

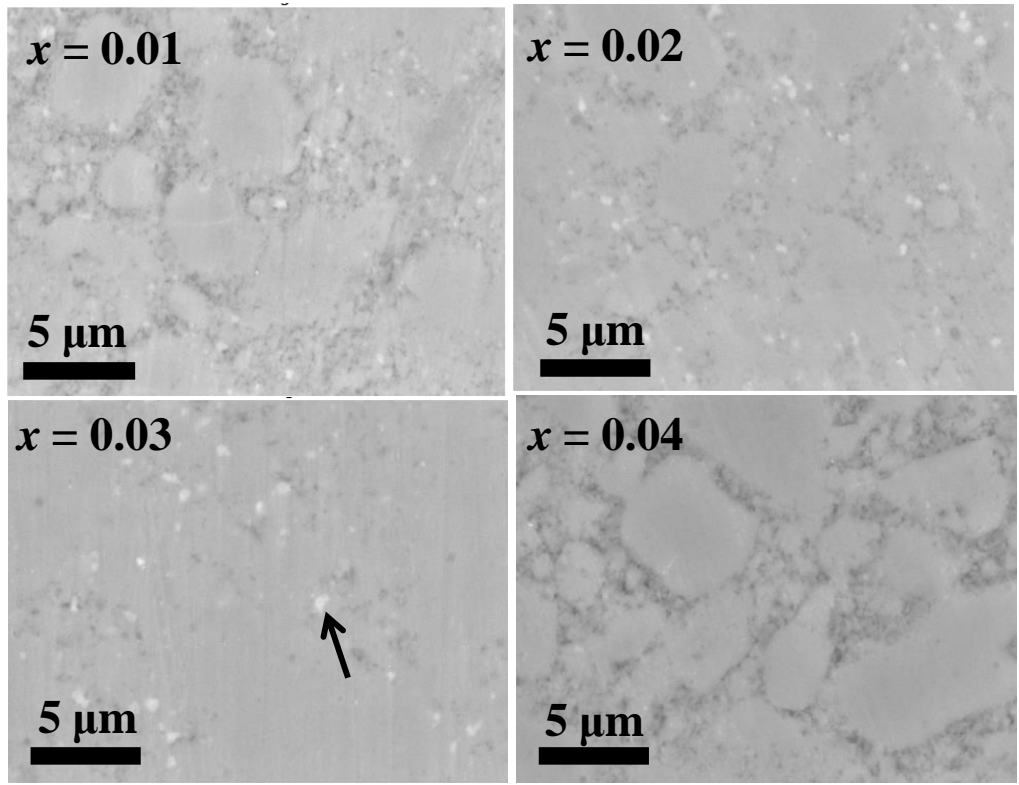

Fig. 5 surface morphology of sintered sample $x = 0.01$, $x = 0.02$, $x = 0.03$, and $x = 0.04$.

*x* = 0.01

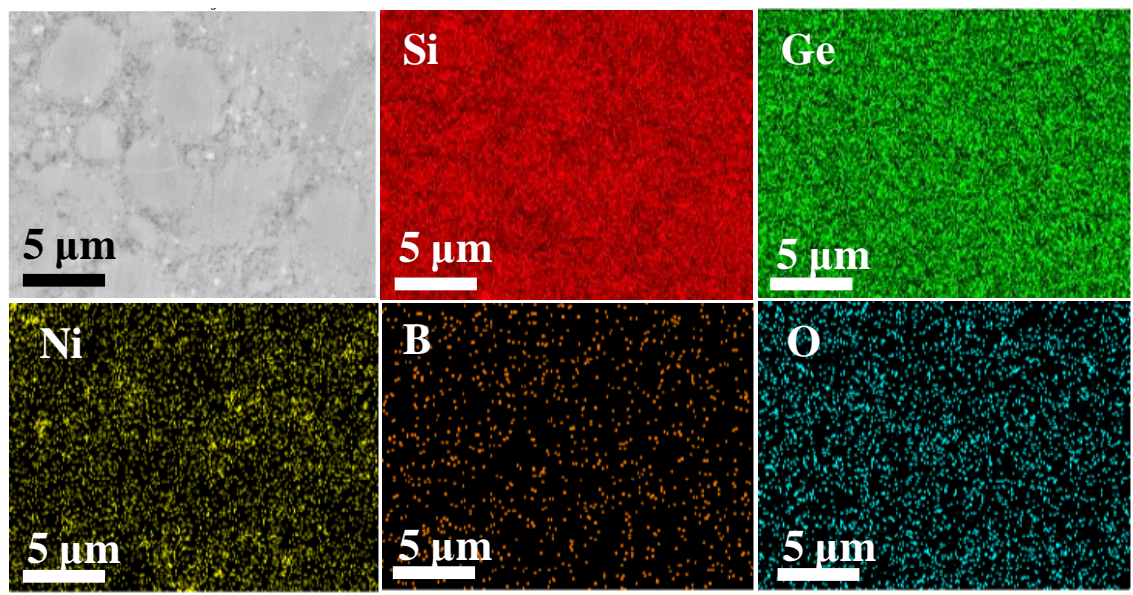

Fig. 6 a

*x* = 0.02

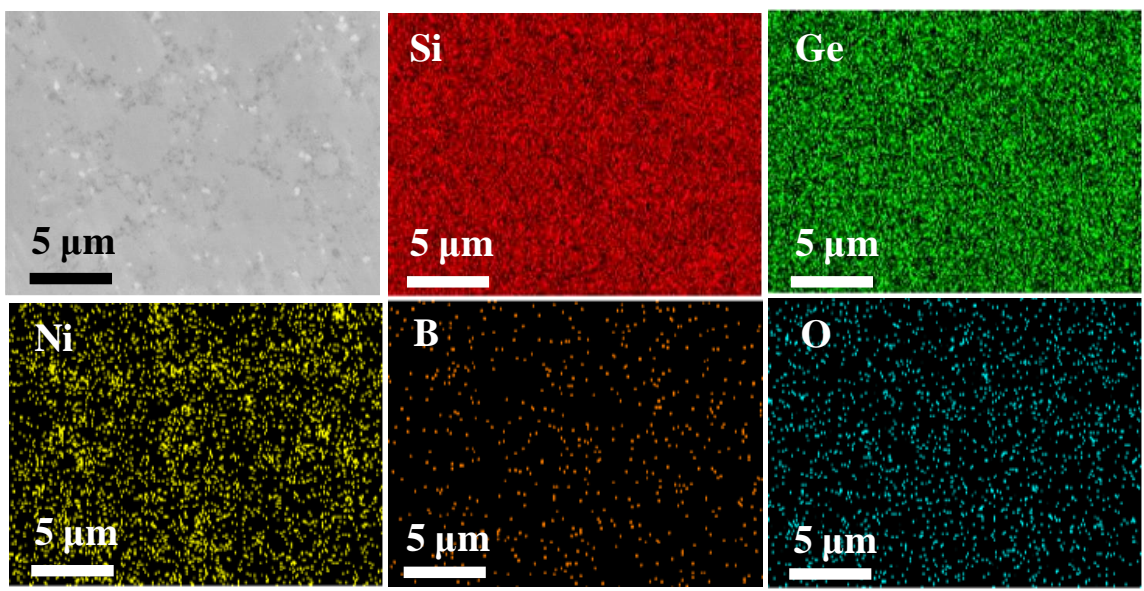

Fig. 6 b

*x* = 0.03

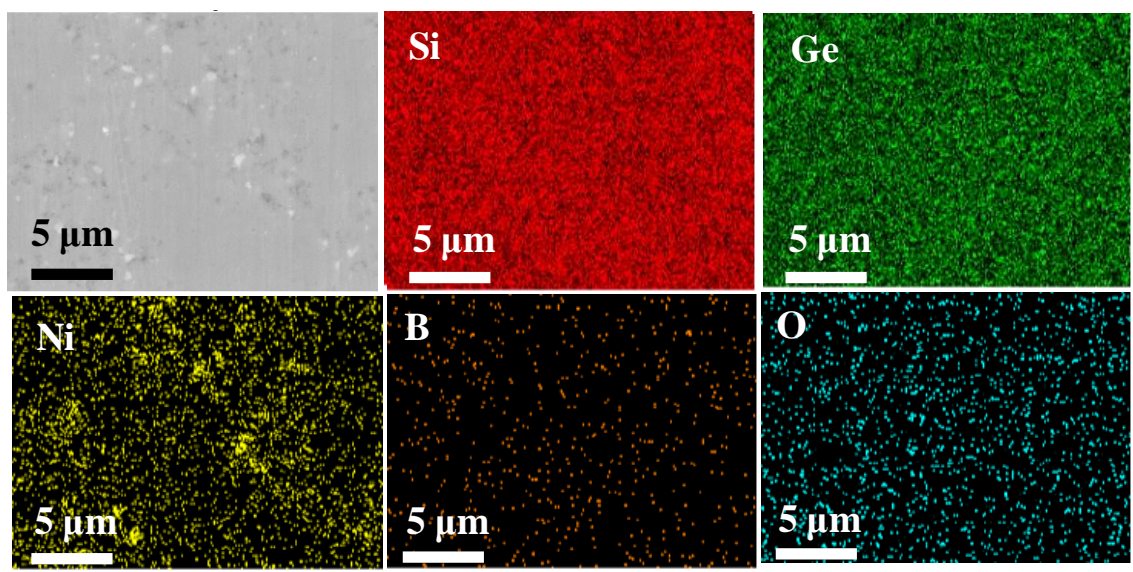

Fig. 6 c

*x* = 0.04

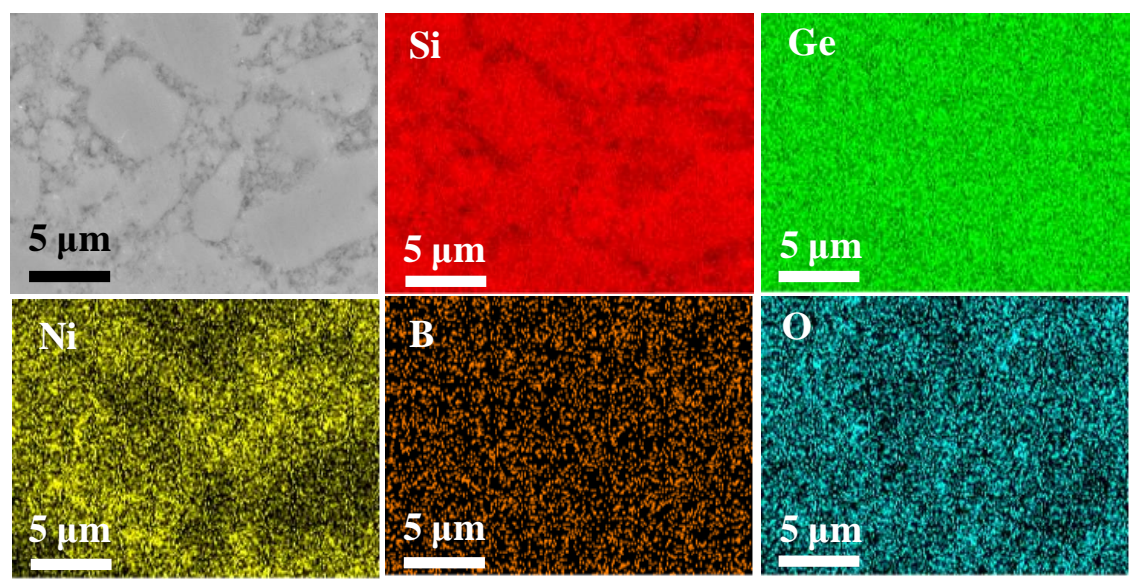

Fig. 6 d

Fig. 6 (a-d) composition mapping analysis of the sample *x* = 0.01 (c) *x* = 0.02, (d) *x* = 0.03, and (e) *x* = 0.04.

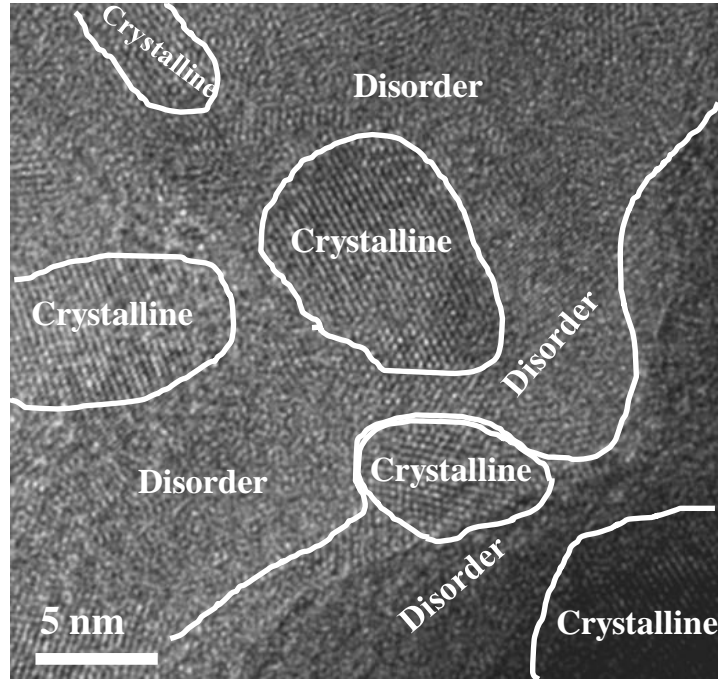

Fig. 7 a

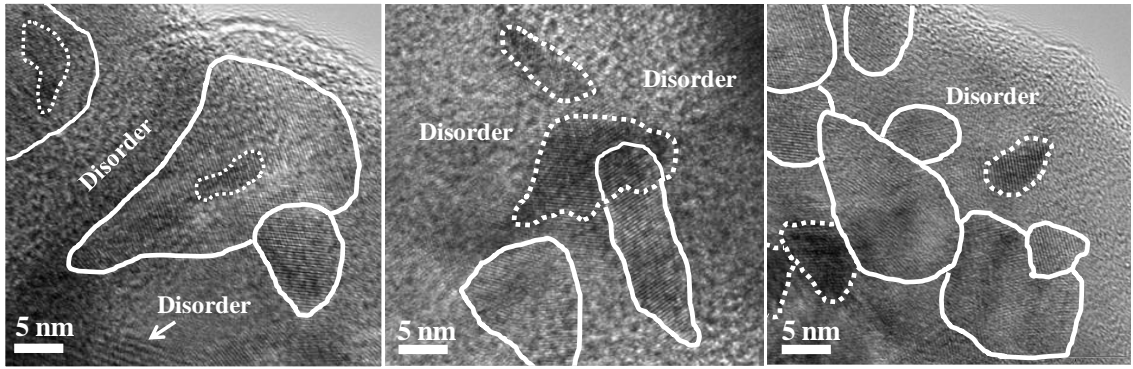

Fig. 7 b

Fig. 7 TEM image of sample $x = 0.03$ (a) after 6 h milled, and (b) crushed sintered sample. The solid and dotted line indicated the Si-Ge nano crystalline and precipitation of Ge-Ni compound. Furthermore, the lattice disorder was also observed for the both 6 h milled and sintered sample.

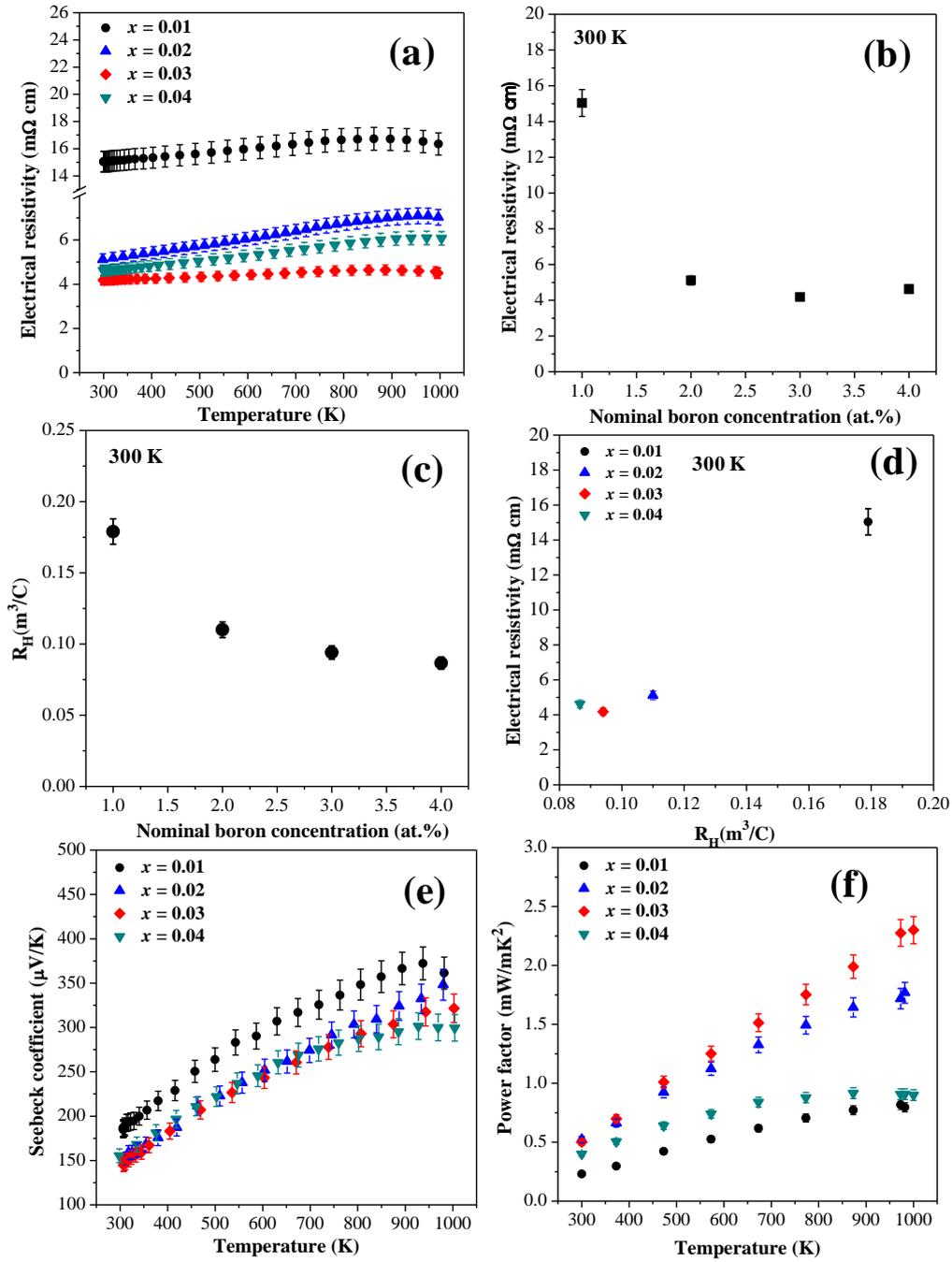

Fig. 8 Thermoelectric properties of the sample $x = 0.01$, $x = 0.02$, $x = 0.03$, and $x = 0.04$. (a) electrical resistivity from 300 K – 1000 K, (b) electrical resistivity at 300 K, (c) Hall coefficient as a function of nominal boron concentration at 300 K, (d) electrical resistivity as a function of Hall coefficient at 300 K, (e) Seebeck coefficient, and (f) power factor from 300 – 1000 K.

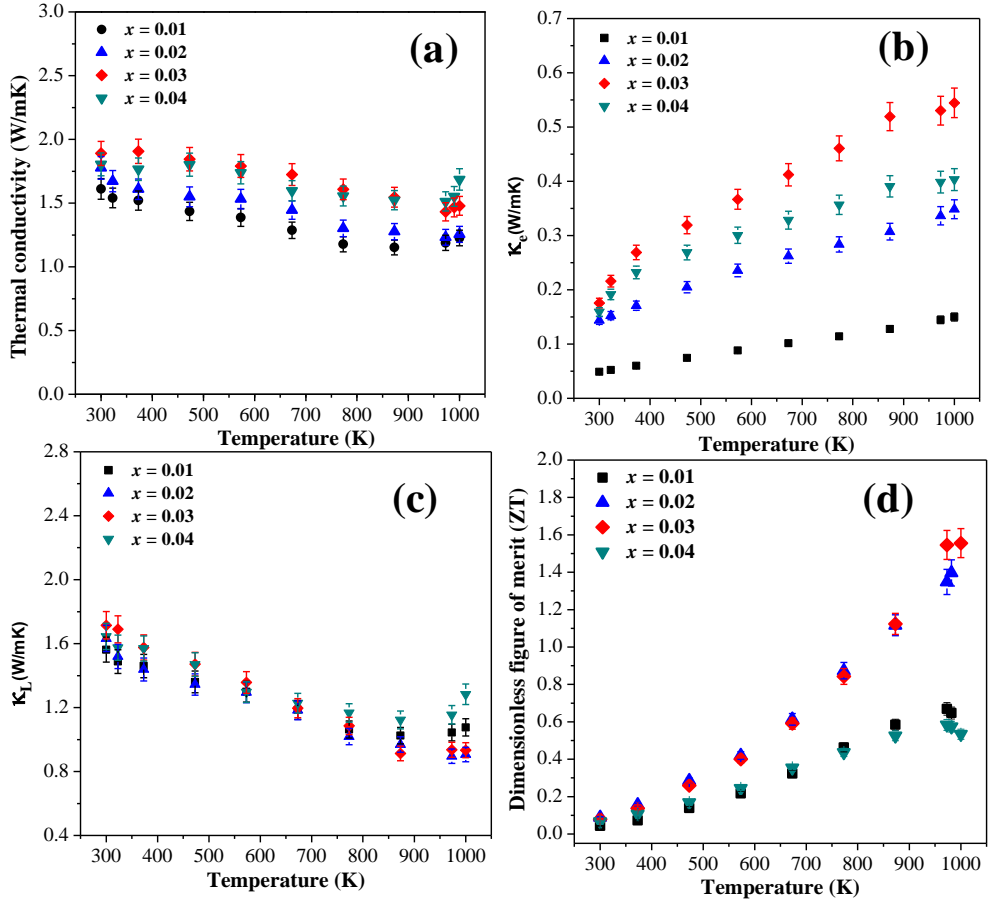

Fig. 9 Thermoelectric properties of the sample $x = 0.01$, $x = 0.02$, $x = 0.03$, and $x = 0.04$ as a function of temperature from 300 – 1000 K (a) total thermal conductivity 'κ', (b) electron thermal conductivity 'κ$_e$', (c) lattice thermal conductivity 'κ$_l$', and (d) dimensionless figure of merit (*ZT*).

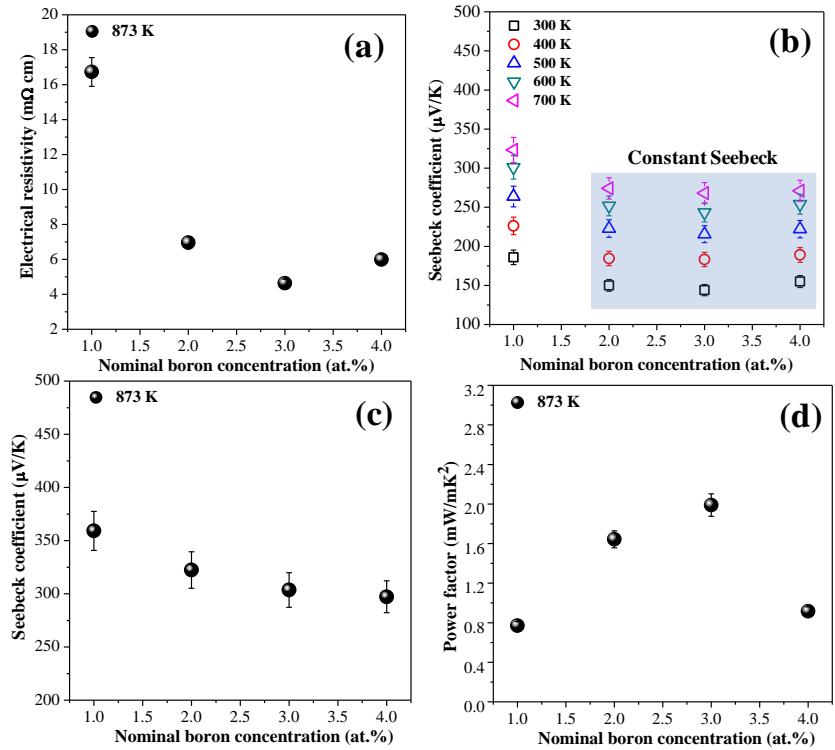

Fig. 10 Thermoelectric properties as a function of nominal boron concentration (a) electrical resistivity at 873 K (b) Seebeck coefficient from 300 K – 700 K, highlighted color indicated the constant Seebeck coefficient was obtained for the sample $x = 0.02$, $x = 0.03$, and $x = 0.04$ (c) Seebeck coefficient at 873 K (d) power factor at 873 K.

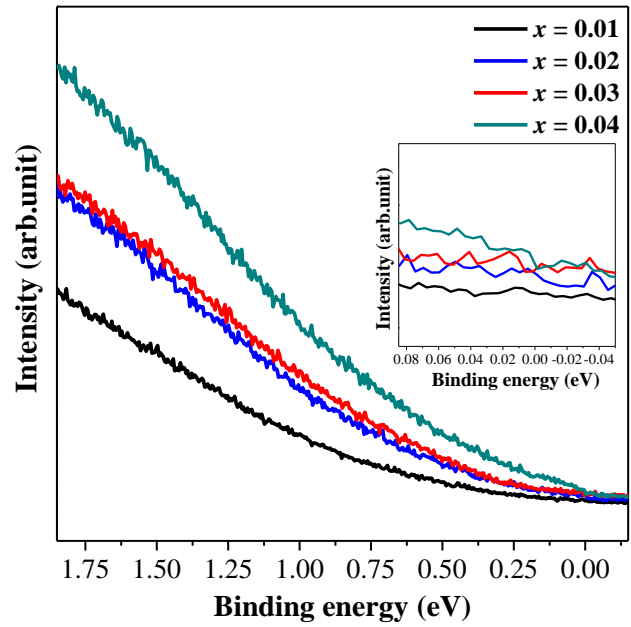

Fig. 11 Photoemission spectroscopy of the sample $Si_{0.62}Ge_{0.35-x}Ni_xB_{0.03}$ ($x$ = 0.01, 0.02, 0.03, and 0.04) at 300 K.

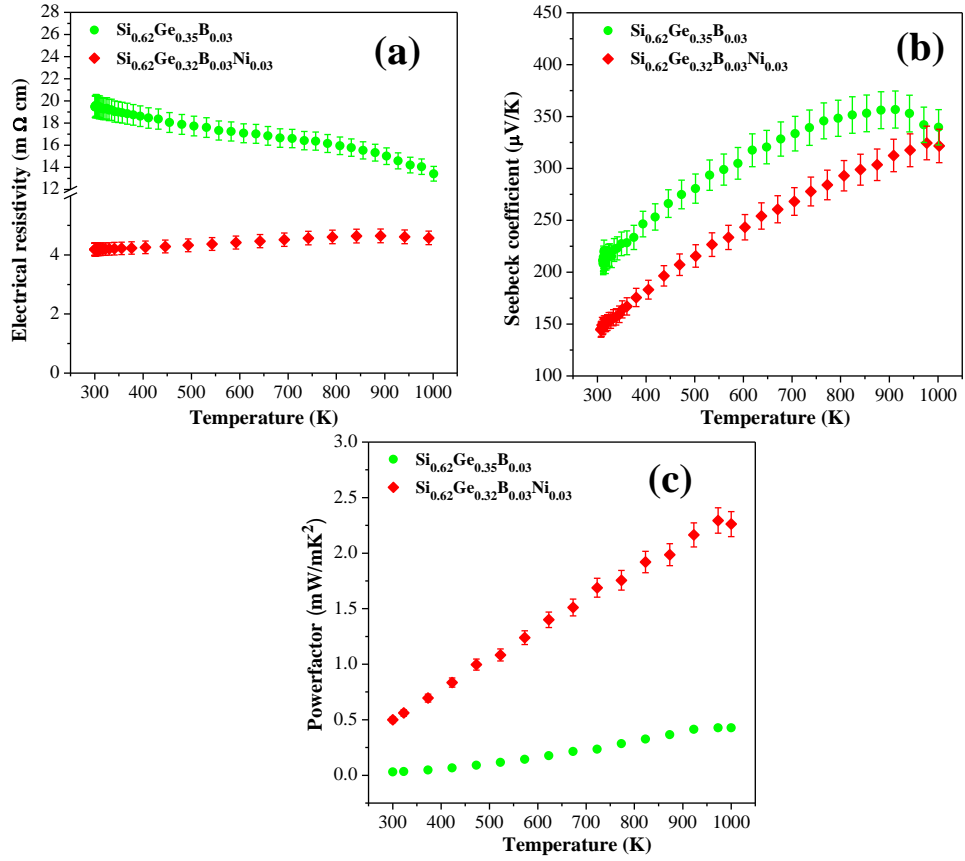

Fig. 12 Thermoelectric properties of the $Si_{0.62}Ge_{0.35}B_{0.03}$, and $Si_{0.62}Ge_{0.32}Ni_{0.03}B_{0.03}$ bulk sample as a function of temperature from 300 K – 1000 K (a) Electrical resistivity (b) Seebeck coefficient (c) Power factor.

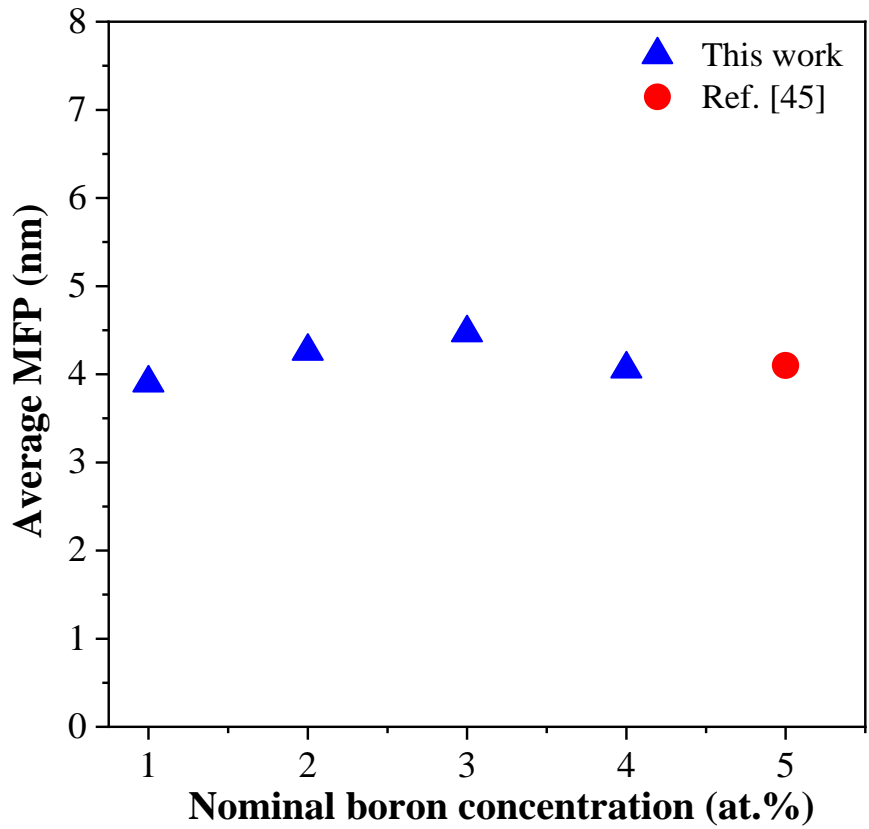

Fig. 13 Average MFP of sample $x = 0.01$, $x = 0.02$, $x = 0.03$, and $x = 0.04$ as a function of nominal boron concentration at 300 K. The red color data point indicates the theoretically estimated MFP in the disorder $Si_{0.8}Ge_{0.2}$ Ref. [45]. The average MFP of all samples was well consistent with theoretically reported value.

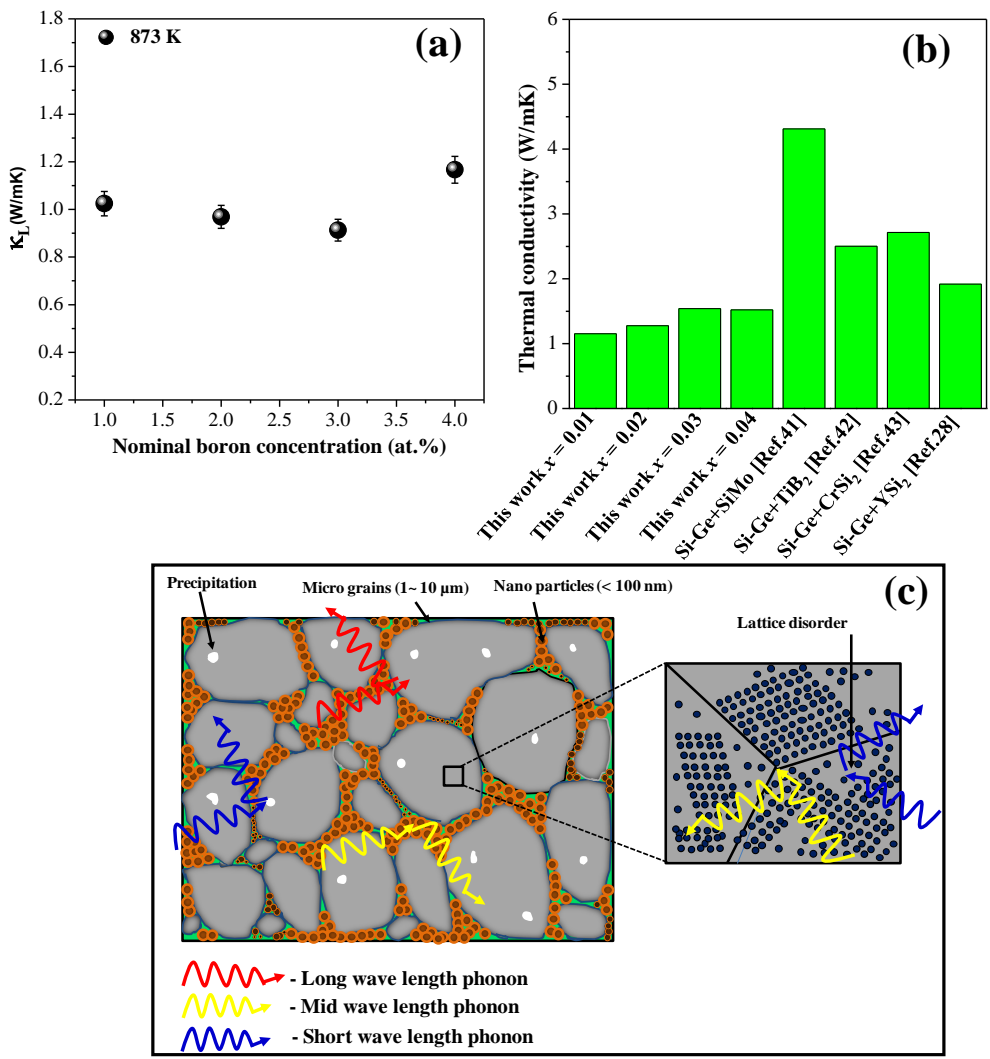

Fig. 14 Thermal conductivity as a function of nominal boron concentration (a) lattice thermal conductivity '$\kappa_l$' at 873 K (b) thermal conductivity of previously reported Si-Ge nanocomposites as compared with this work at 873 K, and (c) schematic view of various wavelength phonons scattering at variable size scattering centers.

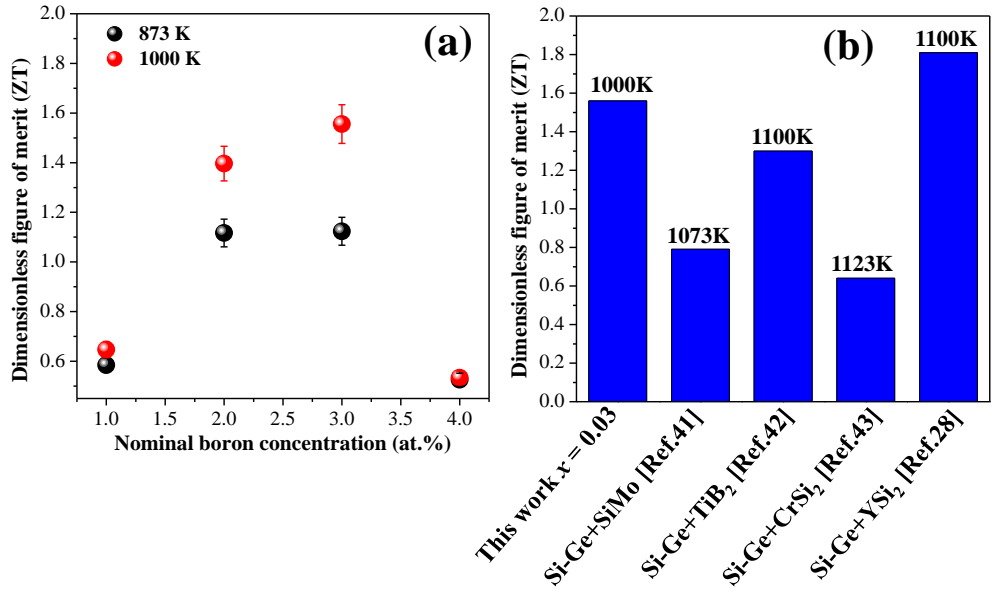

Fig. 15 Dimensionless figure of the sample $x = 0.01$, $x = 0.02$, $x = 0.03$, and $x = 0.04$. as a function of nominal boron concentration (a) 873 K, and 1000 K (b) previously reported Si-Ge nanocomposites as compared with this work at 1000 K.

Table 1: The lattice constant and unit cell volume of the sample $Si_{0.65-x}Ge_{0.32}Ni_{0.03}B_x$ ($x = 0.01$,

0.02, 0.03, and 0.04)

| sample | Lattice constant (Å) | Unit cell volume (Å$^3$) |
|---|---|---|
| $x = 0.01$ | 5.494 | 165.83 |
| $x = 0.02$ | 5.499 | 166.28 |
| $x = 0.03$ | 5.500 | 166.37 |
| $x = 0.04$ | 5.503 | 166.34 |

Table 2: Experimental density compared with theoretical density of samples Si$_{0.65-x}$Ge$_{0.32}$Ni$_{0.03}$B$_x$ ($x = 0.01$, 0.02, 0.03, and 0.04) and crystallite size were calculated from XRD results using Scherr.

| Sample name | Experimental density (g/cm$^3$) | Theoretical density (g/cm$^3$) | Relative density (%) | Crystallite size (nm) |
|---|---|---|---|---|
| $x = 0.01$ | 3.148 | 3.439 | 91.5 | 26 |
| $x = 0.02$ | 3.259 | 3.425 | 95.1 | 27 |
| $x = 0.03$ | 3.284 | 3.412 | 96.2 | 29 |
| $x = 0.04$ | 3.211 | 3.398 | 94.5 | 27 |

Table 3: The composition analysis by electron proble micro analyzer (EPMA) for the

samples $Si_{0.65-x}Ge_{0.32}Ni_{0.03}B_x$ ($x$ = 0.01, 0.02, 0.03, and 0.04) and $Si_{0.62}Ge_{0.35}B_{0.03}$.

| Samples | Si (at.%) | Ge (at.%) | Ni (at.%) | B (at.%) | O (at.%) |
|---|---|---|---|---|---|
| $x$ = 0.01 | 64.6 | 31.23 | 0.27 | 0.68 | 3.13 |
| $x$ = 0.02 | 66.53 | 28.93 | 0.30 | 1.5 | 2.69 |
| $x$ = 0.03 | 65.80 | 29.22 | 0.15 | 2.13 | 2.6 |
| $x$ = 0.04 | 61.9 | 29.8 | 0.26 | 3.2 | 4.6 |
| $Si_{0.62}Ge_{0.35}B_{0.03}$ | 62.7 | 30.4 | 0 | 1.98 | 4.7 |

Table 4: The carrier concentrations and mobility at 300 K

| Samples | Carrier concentration (x $10^{19}$ cm$^{-3}$) | Mobility (cm$^2$/V.s) |
|---|---|---|
| $x$ = 0.01 | 3.49 | 11.9 |
| $x$ = 0.02 | 5.63 | 21.6 |
| $x$ = 0.03 | 6.63 | 23 |
| $x$ = 0.04 | 7.2 | 18.7 |

Table 5: The sound velocity of longitudinal, transverse, and average mean free path of phonon for the sample $Si_{0.65-x}Ge_{0.32}Ni_{0.03}B_x$ ($x$ = 0.01, 0.02, 0.03, and 0.04)

| Sample name | Sound velocity of longitudinal wave (m/s) | Sound velocity of transverse wave (m/s) | Average mean free path of phonon (nm) |
|---|---|---|---|
| $x$ = 0.01 | 6025 | 3692 | 3.9 |
| $x$ = 0.02 | 6800 | 3877 | 4.26 |
| $x$ = 0.03 | 7163 | 4052 | 4.47 |
| $x$ = 0.04 | 6389 | 3749 | 4.06 |
| $Si_{0.8}Ge_{0.2}$ | - | - | 4.1 Ref. [46] |

Table 6: The maximum Seebeck coefficient of $Si_{0.65-x}Ge_{0.32}Ni_{0.03}B_x$ ($x$ = 0.01, 0.02, 0.03, and 0.04) and previously reported $p$-type Si-Ge nanocomposites

| Sample name | Seebeck coefficient ($\mu VK^{-1}$) | Temperature (K) | References |
|---|---|---|---|
| $x$ = 0.01 | 372 | 937 | This work |
| $x$ = 0.02 | 348 | 980 | This work |
| $x$ = 0.03 | 321 | 1000 | This work |
| $x$ = 0.04 | 300 | 985 | This work |
| SiGe+YSi$_2$ | 315 | 1100 | Ref. [28] |
| SiGe+Mo | 180 | 1000 | Ref. [41] |
| SiGe+TiB$_2$ | 350 | 1100 | Ref. [42] |
| SiGe+CrSi$_2$ | 300 | 1000 | Ref. [43] |